\documentclass[aps,twocolumn,superscriptaddress]{revtex4}
\usepackage{xspace,amsmath,amsfonts,amsthm,amssymb,amsbsy,graphics,graphicx,bm}
\usepackage{epigraph}
\usepackage[usenames,dvipsnames]{color}

\usepackage{soul} 
\usepackage{epstopdf}
\usepackage{bbding}
\usepackage{pifont}

\usepackage{tikz}
\usepackage{coffee4}

\usepackage[breaklinks=true]{hyperref}

\hypersetup{
  colorlinks   = true, 
  urlcolor     = blue, 
  linkcolor    = blue, 
  citecolor   = red 
}

\newcommand{\smallfrac}[2]{\mbox{$\frac{#1}{#2}$}}
\newcommand{\half}{\smallfrac{1}{2}}

\newcommand{\bra}[1]{\left\langle{#1}\right|}
\newcommand{\ket}[1]{\left|{#1}\right\rangle}
\newcommand{\op}[2]{\ket{#1}\!\bra{#2}}

\newcommand{\ip}[2]{\left\langle{#1}\right|\left.{#2}\right\rangle}

\newcommand{\erf}[1]{Eq.~(\ref{#1})}
\newcommand{\erfsub}[2]{Eq.~(\ref{#1}\!\! \!#2)}
\newcommand{\frf}[1]{Fig.~\ref{#1}}
\newcommand{\srf}[1]{Sec.~\ref{#1}}

\setstcolor{Magenta}

\newcommand\Id{\mathbb{I}}
\newcommand{\zol}{\mbox{0-1-$\lambda$}}

\begin{document}
\title{Cost of postselection in decision theory} 

\author{Joshua Combes}
\affiliation{Center for Quantum Information and Control, University of New Mexico, Albuquerque, New Mexico, 87131-0001}
\affiliation{Centre for Engineered Quantum Systems, School of Mathematics and Physics, The University of Queensland, St Lucia, QLD 4072, Australia}
\affiliation{Institute for Quantum Computing, University of Waterloo, Ontario N2L 3G1, Canada}
\affiliation{Perimeter Institute for Theoretical Physics, 31 Caroline St. N, Waterloo, Ontario, Canada N2L 2Y5}

\author{Christopher Ferrie}
\affiliation{Center for Quantum Information and Control, University of New Mexico, Albuquerque, New Mexico, 87131-0001}
\affiliation{Centre for Engineered Quantum Systems, School of Physics, The University of Sydney, Sydney, NSW, Australia}


\begin{abstract}
Postselection is the process of discarding outcomes from statistical trials that are not the event one desires.  Postselection can be useful in many applications where the cost of getting the wrong event is implicitly high.  However, unless this cost is specified exactly, one might conclude that discarding all data is optimal.  Here we analyze the optimal decision rules and quantum measurements in a decision theoretic setting where a pre-specified cost is assigned to discarding data. Our scheme interpolates between unambiguous state discrimination (when the cost of postselection is zero) and a minimum error measurement (when the cost of postselection is maximal). We also relate our formulation to previous approaches which focus on minimizing the probability of indecision. 
\end{abstract}

 \date{\today}

\maketitle

\cofeAm{0.25}{0.6}{90}{8in}{-1.5in}

\vspace{-22pt}
\section{Introduction}
There has been some confusion over the role of postselection in quantum information processing protocols. On one hand, postselection is a powerful computational resource~\cite{Aaronson05} and enables technological goals, such as probabilistic photon-photon gates \cite{ObrPryWhi03}. On the other hand, in some situations postselection can impede quantum information processing. 

Probabilistic metrology---also known as metrology with abstention~\cite{GenRonCal2013a} and weak value amplification~\cite{DixStaJor09}---is the idea that postselection may improve estimation precision beyond the usual quantum limits. When the performance of probabilistic metrology is evaluated with respect to the standard figure of merit for parameter estimation, mean squared error, postselection is provably suboptimal, even when there are imperfections~\cite{Knee2013a,TanYam2013,FerCom2013,Knee2013b,ComFerJia13a,ZhaDatWam13,KneComFerGau14,PanJiaCom13}.  Counter claims have been made in the literature (see Refs.~\cite{JorMarHow14,PanDreBru14,CalBenMun14,PangBrun14,SusaTanaka15}) but the issue is far from settled.

In this article we attempt to reconcile the intuition that postselection can help statistical tasks with the fact that for the standard figures of merit generically it does not. To simplify the analysis and make our assumptions explicit we will use a statistical decision theory approach in the context of quantum state discrimination~\cite{BerHerHil04,BarCro09}. To assert that a state discrimination protocol is optimal, we must first specify a {\em cost} or {\em loss} function which encapsulates how each decision is penalized. Then we minimize the average loss over decision rules and measurements. 

This approach defines a task for which the optimal protocol incurs the least losses for the specified loss function. For example consider a two party discrimination game involving an employer Alice and an employee Bob. Alice gives Bob one of two quantum states $\Psi_1$ or $\Psi_2$. Bob is allowed to perform any generalized measurement on the state but then must report which state Alice gave him; he cannot decline to report a state. Bob's bonus, of at most $\mathbb{D} $ dollars, is tied to his performance in this game. If he reports $\Psi_i$ when $\Psi_j$ is true his bonus will be reduced to $\$ (1- \lambda_{i,j} )\mathbb{D}$  where $\lambda_{i,j}$ is called the loss function. Bob wants to devise a strategy to minimise his expected losses. When the cost of reporting the correct answer is ``0" and the incorrect answer is ``1" or maximal, $\lambda_{i,j}$ is known as the $0$-$1$ loss function. Mimimising the losses from the  $0$-$1$ loss function is equivalent to minimizing the probability of misidentifying the states (termed the error probability)~\cite{Helstrom1976Quantum,Fuchs96}. The corresponding optimal measurement strategy, with respect to minimizing losses, is called the Helstrom~\cite{Helstrom1976Quantum} or minimum error measurement. A postselected strategy will have higher expected losses, that is it is suboptimal with respect to the $0$-$1$ loss function.

Postselected strategies for state discrimination were introduced by Ivanovic~\cite{Iva87}, Dieks~\cite{Die88}, and Peres~\cite{Per88} in what is now known as \emph{unambiguous state discrimination} (USD). In USD one allows for an extra ``reject'' decision---postselection---then two nonorthogonal states can be distinguished without error, albeit probabilistically. The USD measurement is optimized in the sense that it has minimal probability of reporting the inconclusive result ``reject''. Prior work on inconclusive state discrimination has focused on exploring and optimizing schemes which interpolate between minimum error probability and minimum inconclusive result probability~\cite{CheBar98,ZhaLiGou99,TouAdaSte07, HayHasHor2008,SugHasHor2009, BagMunOli12, DreBruKor14}. Typically in USD and its generalizations~\cite{CroAndBar06} there is no explicit penalty for reporting ``reject''. 
It is unclear if such postselection is optimal with respect to any loss function. 

Here we re-formalize the inconclusive state discrimination problem by assigning a cost to discarded outcomes.  In particular, we modify the most commonly used cost, the 0-1 loss function, to what we call the \zol\ loss function. In the \zol\ loss function, $\lambda$ is the cost of reporting ``reject''.  In our approach, we find that the USD measurement appears when $\lambda\to 0$. In this limit there is an alternative protocol which is equally optimal: always report ``reject''. Finally we show how our results can be connected to previous approaches where there is a tradeoff between the rejection probablity and the error probablity ~\cite{CheBar98,ZhaLiGou99,TouAdaSte07, HayHasHor2008,SugHasHor2009, BagMunOli12, DreBruKor14}. Our analysis adheres to the desiderata suggested in Ref.~\cite{ComFerJia13a}, and thus is a definitive case where employing postselection can be said to be optimal. 

\cofeAm{0.125}{0.6}{0}{7in}{8in}

\section{Statistical decision theory}
We start by reviewing statistical decision theory and formally introducing the \zol\ loss function, which is a special case of Chow's work on hypothesis testing or classification \cite{Chow57,Chow70}. Consider a set of competing hypotheses $\mathcal H_j$ for $j\in \{1,2,...,n \}$ with prior probabilities $\Pr(\mathcal H_j)$. Given some data $\mathbf{D}$ the posterior probability of the $j$'th hypothesis is 
\begin{align}
\Pr(\mathcal H_j| \mathbf{D}) = \frac{\Pr( \mathbf{D}|\mathcal H_j)\Pr(\mathcal H_j)}{\Pr(\mathbf{D})},
\end{align}
where
\begin{align}
\Pr( \mathbf{D}) = \sum_{j=1}^n \Pr( \mathbf{D}|\mathcal H_j)\Pr(\mathcal H_j).
\end{align}
What we would like to do is have a {\em decision rule} $\delta(\mathbf D)$ that  maps the data $\mathbf{D}$ to decision $i$---that is, report hypothesis $i$, where in this case $i\in \{0,1,2...,n\}$. The decision $i=0$  allows for the possibility that one may not be able to decide, often referred to as the ``don't know'' or ``abstain'' or ``reject" option. 

In Bayesian decision theory the decision rule must arise from minimizing a loss function, which encapsulates how each decision is penalized. The conditional risk, i.e. the {\em a posteriori} expected loss, for the decision  $i$ conditioned on data $\mathbf{D}$ is 
\begin{align}
\mathcal R [i|\mathbf{D}] 
&= \sum_{j=1}^n\   \lambda_{i,j} \Pr( \mathcal H_j| \mathbf{D} ),
\end{align}
where the loss function is denoted by $\lambda_{i,j}$ which corresponds to reporting hypothesis $i$ when hypothesis $j$ is true. The loss function $\lambda_{i,j}$ is a good place to start building intuitions for the role of postselection in detection and estimation theory.

Following Chow, we will require that 
\begin{align}
 \lambda_{i,i} <\lambda_{0,j} <\lambda_{i,j}\quad (i\neq j \neq 0),
\end{align}
which is interpreted as the loss for making a correct decision $ \lambda_{i,i}$ ($i\neq0$) is less than the cost of reject a decision $\lambda_{0,j}$ which is less than the cost of making a wrong decision $\lambda_{i,j}$. We relax this assumption in \srf{beyondzol}, such that  $\lambda_{0,j} >\lambda_{i,j}$ is possible.  A good description of the mathematical and philosophical requirements of a loss function can be found in chapter 2 of Ref.~\cite{berger85}.

The optimal decision is 
\begin{align}
\delta^*(\mathbf{D})  \equiv \arg\min_{i}   \mathcal R [i|\mathbf{D}]. 
\end{align}
When we turn our attention to quantum hypothesis testing we will need to determine the optimal measurement to pair with this optimal decision rule. The criterion for optimal we adopt will require us to minimize the average of the posterior risk
\begin{subequations}\label{total risk}
\begin{align}
\mathcal R[\delta(\mathbf D)] &= \sum_{\mathbf D} \sum_j \lambda_{\delta(\mathbf D),j} \Pr(\mathcal H_j|\mathbf D) \Pr(\mathbf D), \\
& = \sum_{\mathbf D} \sum_j \lambda_{\delta(\mathbf D),j} \Pr(\mathbf D|\mathcal H_j) \Pr(\mathcal H_j),
\end{align}
\end{subequations}
over the distribution of data and the measurement. When we assume the optimal decision is being used we denote the total risk as $\mathcal R ^*=\mathcal R[\delta^*(\mathbf D)] $.

To simplify or analysis we will consider binary hypothesis testing (i.e $\mathcal H_1$ vs $\mathcal H_2$) and take
\begin{align}
\begin{array}{rl}\label{01lam}
\lambda_{1,1} &=\lambda_{2,2} = 0,\\
\lambda_{1,2} &=\lambda_{2,1} =1,\\
\lambda_{0,1}&=\lambda_{0,2}=\lambda,
\end{array}
\end{align}
which we call the ``\zol'' loss function. For the \zol\ loss function the conditional risks for decisions $i$ are
 \begin{align}\label{cond_risky}
 \begin{array}{rl}
\mathcal R [2|\mathbf{D}] 
&= 1-\Pr( \mathcal H_2| \mathbf{D} ),\\
\mathcal R [1|\mathbf{D}]
&=1-\Pr( \mathcal H_1| \mathbf{D} ),     \\
\mathcal R [0|\mathbf{D}] 
&= \lambda   ,
\end{array}
\end{align}
where we have used $\Pr( \mathcal H_1| \mathbf{D} )+ \Pr( \mathcal H_2| \mathbf{D} )=1$.

\begin{figure}\centering
  \includegraphics[width=\columnwidth]{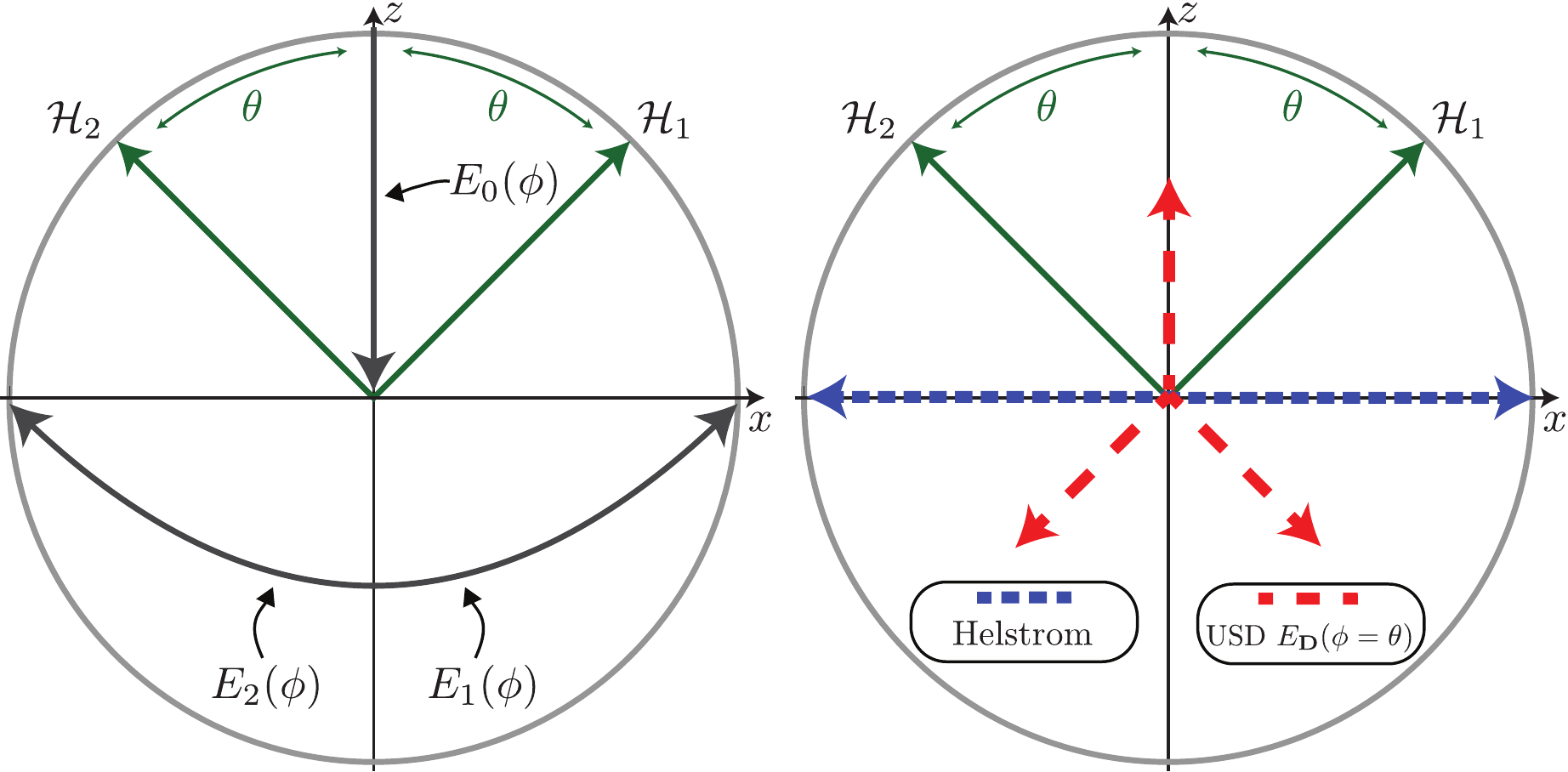}
  \caption{\label{fig}  The Bloch representation of the states and POVM elements involved in the state discrimination protocol.  The POVM elements, $E_{\bf D}(\phi)$ are not mixed states, but subnormalized rank-1 operators, which lie on a circle at a lower level in a cone of positive operators. The grey lines on the left figure are the arc of the POVM elements as $\phi$ is varied in \erf{POVM} from 0 to $\pi/2$. The right figure is illustrates two special cases of the POVM elements $E_{\bf D}(\phi)$. When $\phi =\pi/2 $ there are only two POVM elements and the measurement is the Helstrom measurement. When $\phi=\theta$ we recover the USD measurement.  }
\end{figure}

Thus our decision rule $\delta^*(\mathbf{D})$  is 
\begin{align}\label{decision1}
\delta^*(\mathbf{D}) =  \begin{cases}
2 & \text{if } \mathcal R [2|\mathbf{D}]  < \mathcal R [1|\mathbf{D}]  \text{ and } \mathcal R [0|\mathbf{D}] \\
1 & \text{if } \mathcal R [1|\mathbf{D}]  < \mathcal R [2|\mathbf{D}]  \text{ and } \mathcal R [0|\mathbf{D}] \\
0 & \text{otherwise} 
\end{cases}.
\end{align}
With respect to the posterior probabilities we find 
\begin{align}\label{decision2}
\delta^*(\mathbf{D}) =  \begin{cases}
2 & \text{if }\Pr( \mathcal H_2| \mathbf{D} )\ge 1- \lambda \text{ and }
   \Pr( \mathcal H_1| \mathbf{D} ) \\
1 & \text{if }\Pr( \mathcal H_1| \mathbf{D} )\ge 1- \lambda \text{ and } 
 \Pr( \mathcal H_2| \mathbf{D} ) \\
0 & \text{otherwise} 
\end{cases}.
\end{align}
In words, the decision rule is as follows: find the largest posterior probability; if it is greater than or equal to the threshold $1-\lambda$, report it; if it is less than $1-\lambda$ report ``reject''. Now we connect this decision theoretic framework to quantum hypothesis testing.

\cofeAm{0.05}{0.6}{90}{8in}{-1.5in}

\section{State discrimination}\label{sec:statediscrim}
In quantum theory the statistics of measurements are described by a positive operator valued measure (POVM) $\{E_{\mathbf D} \}$, the elements of which sum to the identity: $\sum_{\mathbf D}E_{\mathbf D}=\Id$ . The number of elements of a POVM is the number of outcomes of the measurement. To match this with our previous terminology the outcomes of the measurement are the data $\mathbf{D}$. In order to encompass both USD and Helstrom measurements we must consider a three-outcome POVM $E_{\mathbf D}$ where $\mathbf D \in \{0, 1, 2\}$.  Let us make the following symmetry assumptions to make the discussion less cumbersome:
\begin{subequations}\label{symm}
\begin{align}
\Pr( \mathcal H_1 )& = \Pr( \mathcal H_2),\\
\Pr(E_1) &=\Pr(E_2),\\
\Pr( E_1| \mathcal H_1) & =\Pr( E_2| \mathcal H_2),\\
\Pr( E_1| \mathcal H_2) & =\Pr( E_2| \mathcal H_1),\\
\Pr( E_0| \mathcal H_1) & =\Pr( E_0| \mathcal H_2).
\end{align}
\end{subequations}
These symmetries are implied, for example, by the states and operators in Fig.~\ref{fig}.

\begin{figure*}\centering
  \includegraphics[width=\textwidth]{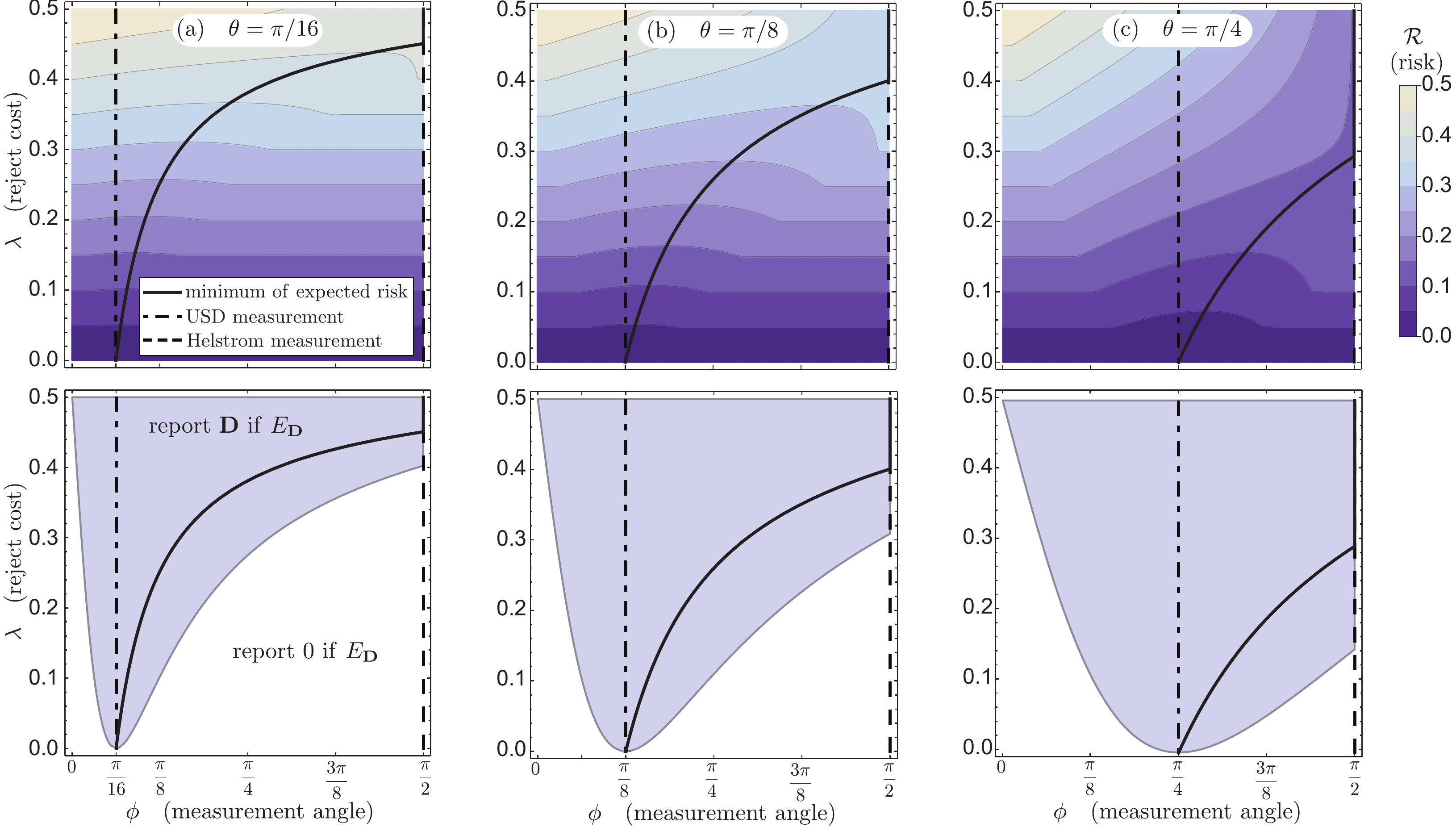}
\caption{\label{fig2} Expected risk $\mathcal{R}$ (row 1)  and decision rule (row 2) for the $\zol$ loss function.  In all figures the abscissa is $\phi$ (the measurement angle)  and  the ordinate is $\lambda$ (the cost of reporting ``reject'' ). The dark black line is the minimum risk ($\mathcal{R}^*[\phi^*]$) for a given $\lambda$ and thus specifies the optimal measurement angle. The shaded regions in the second row are simply the region for which the expected risk is less than $\lambda$; in this region one always reports $i$ if one obtained outcome $E_i$.}
\end{figure*}

Utilizing some of these these symmetries the total risk in \erfsub{total risk}{\,b} becomes
\begin{align}
\mathcal R 
=& \half [ (\lambda_{\delta(0),1}+\lambda_{\delta(0),2})\Pr(E_0|\mathcal H_1)+\nonumber\\
&\quad (\lambda_{\delta(1),1}+\lambda_{\delta(2),2})\Pr(E_1|\mathcal H_1)+\nonumber\\
&\quad (\lambda_{\delta(2),1}+ \lambda_{\delta(1),2})\Pr(E_2|\mathcal H_1)].
\end{align}
 Next we use the optimal decision rule, \erf{decision1} or \erf{decision2}, and more of the symmetries to massage this expression. Further, we assume that $\lambda <1/2$; as for $\lambda\ge1/2$ one can always randomly choose to report $\mathcal H_1$ or $\mathcal H_2$ and reduce the expected risk (in \srf{beyondzol} we will relax this assumption). Equation (\ref{symm}{e})  implies $\Pr(\mathcal H_1|E_0)=\Pr(\mathcal H_2|E_0)=1/2$, thus the lowest conditional risk i.e. \erf{cond_risky} implies that the optimal decision for $\mathbf D=0$ is $\delta^*(0)=0$ always. Also $\lambda_{{\delta^*}(1),1}=\lambda_{\delta^*(2),2}$ and $\lambda_{\delta^*(2),1}=\lambda_{\delta^*(1),2}$ are implied by symmetry as well.  Using these relations we obtain
\begin{align}
\mathcal R^* 
=& \lambda_{0,1} \Pr(E_0|\mathcal H_1)+ \lambda_{\delta^*(1),1}\Pr(E_1|\mathcal H_1)+\nonumber\\
& \lambda_{\delta^*(2),1}\Pr(E_2|\mathcal H_1).
\end{align}
Recall from \erf{01lam} that $\lambda_{0,1}= \lambda$. Using this and Bayes rule we obtain
\begin{align}
\mathcal R^* =&2[ \lambda\Pr(\mathcal H_1|E_0) \Pr(E_0)+\lambda_{\delta^*(1),1}\Pr(\mathcal H_1|E_1)\Pr(E_1)\nonumber\\
&+\lambda_{\delta^*(2),1}\Pr(\mathcal H_1|E_2)\Pr(E_2)].
\end{align}
Then using $\Pr(E_0)= 1-\Pr(E_1)-\Pr(E_2)=1-2\Pr(E_1)$ we have
\begin{align}
\mathcal R^* =&2\{  \half \lambda[1-2 \Pr(E_1)]+\\
& [\lambda_{\delta^*(1),1}\Pr(\mathcal H_1|E_1)+\lambda_{\delta^*(1),2}\Pr(\mathcal H_2|E_1)]\Pr(E_1)\},\nonumber
\end{align}
where we have used $\Pr(\mathcal H_1|E_2) = \Pr(\mathcal H_2| E_1)$ and \erfsub{symm}{b}.
The term $T=[\lambda_{\delta^*(1),1}\Pr(\mathcal H_1|E_1)+\lambda_{\delta^*(2),1}\Pr(\mathcal H_1|E_2)]$ still depends on the optimal decision rule so we must explictly use it. It is important to note that we can't assume $\delta^*(1)=1$, this means we must consider two cases ($\delta^*(1)=2$ is obviously ruled out by symmetry): (1) $\delta^*(1)=0$: this implies $T= \lambda [ \Pr(\mathcal H_1|E_1)+ \Pr(\mathcal H_2|E_1)]=\lambda$; or (2) $\delta(1)=1$: this implies $T= \Pr(\mathcal H_2|E_1)$.
Using the optimal decision rule, the risk becomes
\begin{align}\label{risk1}
\mathcal R^* &= \begin{cases}
\lambda\Pr(E_0|\mathcal H_2)+\Pr(E_1|\mathcal H_2) & \text{if } \Pr(\mathcal H_2|E_1)\leq \lambda\\
\lambda & \text{otherwise}
\end{cases}.
\end{align}
Equivalently this can be written as
\begin{align}\label{da risk}
\mathcal R^* &= \lambda + \min \left \{ 0,\Pr(E_1|\mathcal H_2)-\lambda [1-\Pr(E_0|\mathcal H_2)] \right \}
\end{align}
The above risk is true for the \zol\ loss function and any two hypotheses and measurements satisfying the symmetry conditions. The first term represents the part of the expected risk when a rejection is made. The second term is not yet optimized over the possible measurements.  

As a specific example, here we will consider the problem of discriminating the following two quantum states:
\begin{subequations}
\begin{align}\label{quantdis}
\mathcal H_1: \quad\ket{\Psi_1} &=  \cos\smallfrac{\theta}{2} \ket{0} + \sin\smallfrac{\theta}{2} \ket{1},\\
\mathcal H_2: \quad\ket{\Psi_2}  &= \cos\smallfrac{\theta}{2} \ket{0} -\sin\smallfrac{\theta}{2} \ket{1},
\end{align}
\end{subequations}
where $0\le \theta \le \pi/2$, $|\ip{\Psi_2}{\Psi_1}|=\cos\theta$ and the prior probabilities are $\Pr(\mathcal H_1)= \Pr(\mathcal H_2)=1/2$.

The symmetry we imposed in \erf{symm}, imply the measurement is in fact a generalized measurement with POVM elements
\begin{align}\label{POVM}
E_2(\phi)&=\frac{1}{{2 \cos ^2\smallfrac{\phi}{2} }}\left(
\begin{array}{cc}
 \sin ^2\smallfrac{\phi}{2}  & -\sin \smallfrac{\phi}{2}  \cos \smallfrac{\phi}{2}  \\
 -\sin \smallfrac{\phi}{2}  \cos \smallfrac{\phi}{2}  & \cos ^2\smallfrac{\phi}{2}  \\
\end{array}
\right),\nonumber\\
E_1(\phi)&=\frac{1}{{2 \cos ^2\smallfrac{\phi}{2} }}\left(
\begin{array}{cc}
 \sin ^2\smallfrac{\phi}{2}  &\phantom{-} \sin \smallfrac{\phi}{2}  \cos \smallfrac{\phi}{2}  \\
 \phantom{-}\sin \smallfrac{\phi}{2}  \cos \smallfrac{\phi}{2}  & \cos ^2\smallfrac{\phi}{2}  \\
\end{array}
\right),\\
E_0(\phi)&=\left(
\begin{array}{cc}
 1-\tan ^2\smallfrac{\phi}{2}  & 0 \\
 0 & 0 \\
\end{array}
\right),\nonumber
\end{align}
such that $E _2(\phi)+E _1(\phi)+E _0(\phi)=\Id$. When $\phi=\pi/2$ we get $E_0= 0, E_1= \op{+}{+}, E_2= \op{-}{-}$ (where $\ket{\pm}$ are the eigenstates of the Pauli $X$ operator), which is the Helstrom measurement for all $\theta$. When $\phi = \theta$ we obtain the USD measurement for all $\theta$. In \frf{fig} the grey lines are the arc traced by \erf{POVM} as a function of $\phi$. Note that for $\phi > \pi/2$ the POVM element $E_0$ is not a positive operator, thus we do not allow these values of $\phi$.

To apply the above decision theoretic formalism we need to compute the probabilities given in  \erf{da risk}. All of these probabilities can be computed using the usual rule: 
\begin{align}
\Pr(E_{\mathbf D}|\mathcal H_i,\phi) = \bra{\Psi_i}E_{\mathbf D}(\phi)\ket{\Psi_i},
\end{align}
see footnote \cite{explicit} for some examples. Notice how all of the probabilities depend on the measurement angle $\phi$, this means the expected risk will also be a function of $\phi$.

Given the POVM elements in \erf{POVM} the expected risk is
\begin{align}\label{risk_eg}
\mathcal{R}^*[\phi]=&\, \lambda\,+\\
&\min \left[0,\frac{(2 \lambda -1) (\cos \theta  \cos \phi -1)-\sin \theta \sin \phi}{2 (1+\cos \phi)}\right],\nonumber
\end{align}
Intuitively this says the risk is at most $\lambda$ and sometimes less. This risk is plotted in \frf{fig2} as a function of $\lambda$ and $\phi$ for particular values of $\theta$. To find the optimal angle we fix $\lambda$ and ask which $\phi$ minimizes  $\mathcal{R}^*[\phi]$. This can be done analytically. The trival case is when $\mathcal{R}^*[\phi]=\lambda$ an thus no optimization over $\phi$ is possible. The optimal measurement found by solving
\begin{align}\label{risk_eg}
\frac{\partial}{\partial \phi}\left [ \lambda\,+\frac{(2 \lambda -1) (\cos \theta  \cos \phi -1)-\sin \theta \sin \phi}{2 (1+\cos \phi)} \right ]=0,
\end{align} 
for $\phi$. { The constraint on the positivity of the measurement operators, i.e. $\phi\le \pi/2$, results in following peicewise defintion of optimal measurement angle} 
\begin{align}\label{phi_opt}
\phi^* = \!\left\{\!
 \begin{array}{cl}
2 \cot^{-1}\!\left[(1-2 \lambda ) \cot \dfrac{\theta }{2}\right] \!& \text{if }\,\lambda < \dfrac 1 2  \left(1 - \tan\dfrac \theta 2  \right )\vspace{5pt}\\
  \dfrac{\pi}{2} &  \text{if }\,\lambda \ge \dfrac 1 2  \left(1 - \tan\dfrac \theta 2  \right )\vspace{5pt}\\
 \end{array} \right ..
\end{align}
This optimal angle is plotted as the solid black lines in \frf{fig2}. The decision functions plotted in the second row of \frf{fig2} are particularly simple: in the shaded regions report $\mathbf D$ if $E_{\mathbf D}$ is observed and report ``reject'' or 0 if $E_{\mathbf D}$ is observed in the non shaded regions.

From \frf{fig2} it is clear that, as a function of $\lambda$ the optimal measurement angle interpolates between the USD and the Helstrom measurement. This can be made explicit as follows. The second branch of  \erf{phi_opt}, i.e. when $\phi^*= \pi/2$, is the Helstrom measurement. To recover the USD measurement we plug $\lambda = 0$ into \erf{phi_opt} gives $\phi^*= \theta$, so $\lambda=0$ implies the USD measurement.  However, $\lambda=0$ is also a degenerate case where no cost is assigned to reporting ``reject''.  Thus, the risk is {\em also} minimized by reporting ``reject'' for \emph{any} outcome of \emph{any} measurement or, equivelently, not bothering to make the measurement and simply reporting ``reject''. Recall that what we are calling \emph{the} USD measurement is the one which minimizes the probability of obtaining the ``reject'' outcome in the usual paradigm. Here, as expected, the USD measurement is approached for $\lambda\to0$. This is also when the probablity for reporting ``reject'' is maximized, see \frf{fig5} of \srf{sec:pr_pe}. 

\begin{figure}\centering
  \includegraphics[width=\columnwidth]{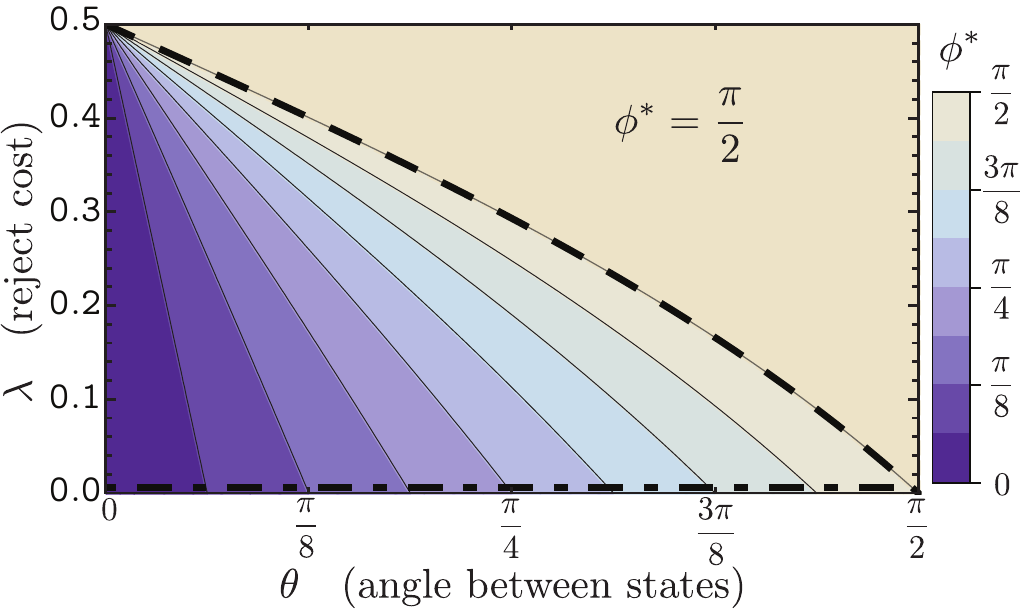}
  \caption{\label{fig3} The angle $\phi^*$ of the optimal measurement minimizing the risk for the $\zol$ loss, i.e. \erf{phi_opt}, as a function of $\lambda$ and $\theta$. The dot dashed line at $\lambda=0$ corresponds to the USD measurement when $\phi^*=\theta$. Above the dashed line the Helstrom measurement is optimal. The optimal angle has been discretized for ploting}.
\end{figure}

To complete the example we plot in \frf{fig3} the optimal measurement angle $\phi^*$ as a function of $\lambda$ and the angle between the states $\theta$ and the $z$ axis. The USD protocol corresponds the line at $\lambda =0$ while the Helstrom measurement is performed for when $\phi^*=\pi/2$. The area where $\phi^*=\pi/2$ is approximately half of the parameter space, i.e. $\lambda \gtrsim \half (1-\theta/2) + O(\theta^3)$, thus even when the loss function encourages postselection it is not guaranteed to be optimal.

Other studies of inconclusive state discrimination \cite{TouAdaSte07,HayHasHor2008, BagMunOli12, DreBruKor14} concern themselves with the probabilities of error and reporting the ``reject'' result.  This avoids the question of what to do \emph{given} the outcome of some measurement.  Here we have phrased the problem as a decision theoretic one where the loss is incurred on the decisions and once that loss is specified, a definitive answer can be given.  In real applications, it would be unlikely that an agent's decisions are constrained to be deterministic functions of measurement operators.  Indeed, our results imply that loosening that constraint can only decrease the agent's risk if they can not measure at the optimal angle for a given $\lambda$.

\section{Relationship between Risk and error and reject probabilities}\label{sec:pr_pe}
So far we have focused on the decision function and the loss function. In this section we connect our approach to the previous approaches which focus on tradeoffs between reject and error probabilities \cite{DreBruKor14}, and rejection thresholds \cite{BagMunOli12}. 

For equal prior probabilities the optimal decision rule when measuring at the optimal angle, is particularly simple: report $\mathbf D$ if $E_{\mathbf{D}}$. Let probability of making the correct decision be $C$, the probability of error be $E$, the probability of rejection be $R$ and the probability that a piece of data is accepted be $A$.  These probabilities can be written explicitly as follows:
\begin{subequations}\label{Probz}
\begin{align}
\Pr(C|\theta,\lambda)&=\sum_{i\in\{1,2\}}\Pr(\mathcal H_i) \Pr[E_i(\phi^*)|\Psi_i],\\
\Pr(E|\theta,\lambda)&=\sum_{i, j\in\{1,2\},i\neq j}\Pr(\mathcal H_i) \Pr[E_j(\phi^*)|\Psi_i],\\
\Pr(R|\theta,\lambda)&=\sum_{i\in\{1,2\}}\Pr(\mathcal H_i) \Pr[E_0(\phi^*)|\Psi_i],\\
\Pr(A|\theta,\lambda)&=\Pr(C|\theta,\lambda) +\Pr(E|\theta,\lambda),
\end{align}
\end{subequations}
These probabilities obey $\Pr(E) +\Pr(C)+\Pr(R)=1$ which implies $\Pr(A)+\Pr(R) = 1$. 

\begin{figure}\centering
  \includegraphics[width=0.95\columnwidth]{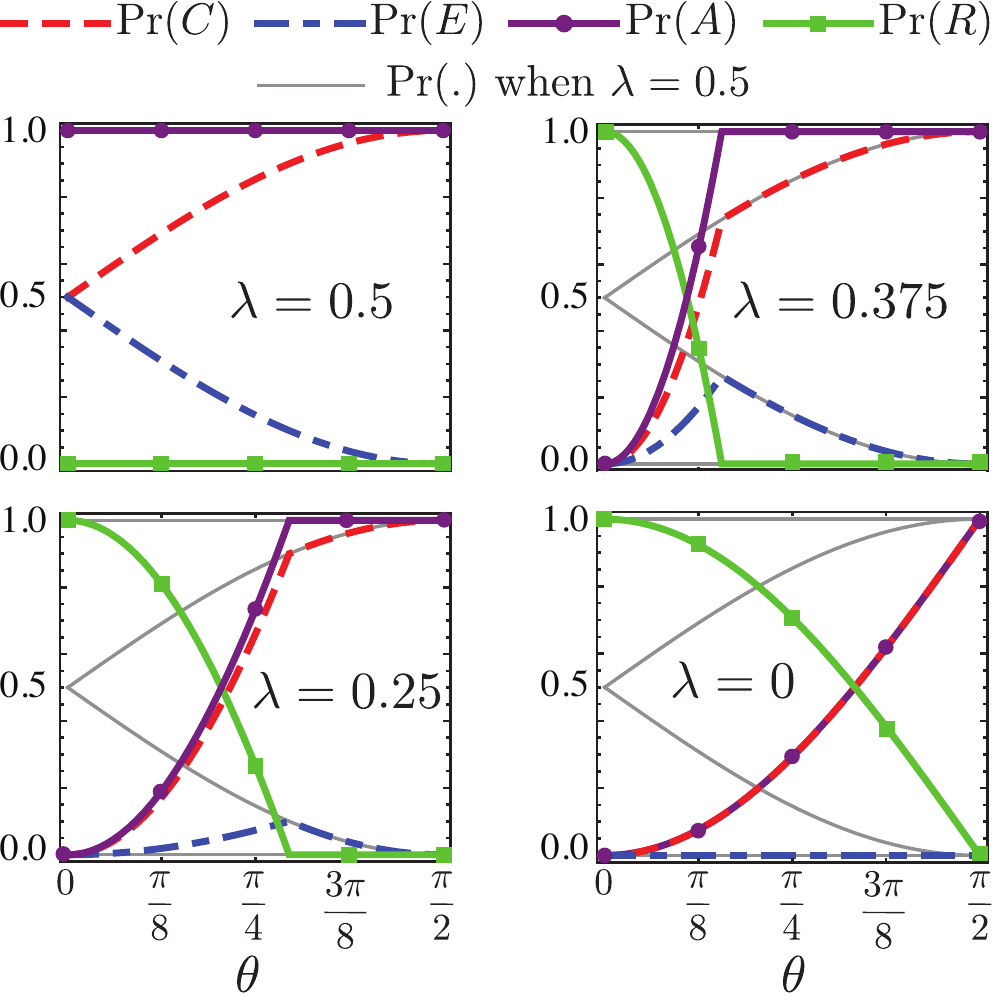}
  \caption{\label{fig4} The probabilities in \erf{Probz} as a function of the angle between the states $\theta$. When $\lambda = 0.5$ it is easy to show that $\Pr(C|\theta)=1-\Pr(E|\theta)=(1/2)(1+\sin\theta)$, $\Pr(A|\theta)= 1$, and $\Pr(R)= 0$ as plotted in the top left plot. These lines are the gray lines on the other figures. Generically as $\theta\rightarrow 0$ probability for reporting ``don't know" indeed approaches one except when $\lambda =0.5$. When the equality in the second branch of \erf{phi_opt} is satisfied we see the measurement switches from one with an inconclusive outcome to the Helstrom measurement i.e. $\Pr(A)=1$ and $\Pr(R)=0$ and $\Pr(C|\theta)=1-\Pr(E|\theta)=(1/2)(1+\sin\theta)$.
}
\end{figure}

In \frf{fig4} we plot these probabilities as a function of the angle $\theta$ between the states. A strategy without postselection adheres to the lines of \frf{fig4} when $\lambda=0$. Deviating from this behavior indicates postselection. Notice that as $\theta\rightarrow 0$ $\Pr(R)\rightarrow 1$ for all $\lambda$ except $\lambda=0.5$. While, in \frf{fig5} we plot the error probability and reject probability as a function of the rejection threshold. Postselection occurs whenever $\Pr(A)<1$. Notice that as $\lambda$ approaches 0, the probablilty of rejection gets closer to 1 for all values of $\theta$.

\begin{figure}\centering
  \includegraphics[width=0.95\columnwidth]{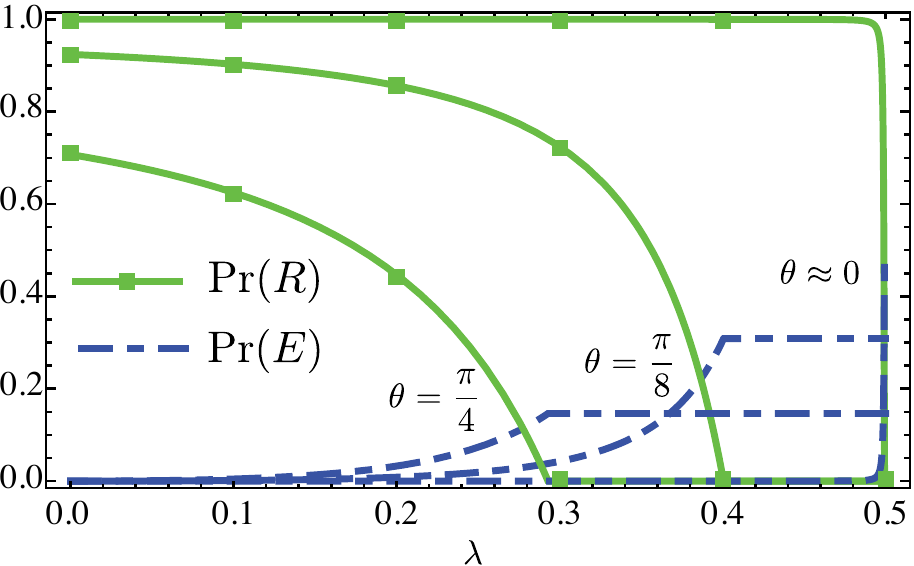}
  \caption{\label{fig5} The rejection and error probabilities as a function of $\lambda$. When $\lambda = 0$ the measurement strategy is precisely the USD measurement and the rejection probability attains its maximum  $\Pr(R)=\cos\theta$. Now consider the values of $\lambda$ for which $\Pr(R)=0$. For example when $\theta = \pi/8$, $\Pr(R)=0$ when $\lambda \in [0.4,0.5]$. As $\lambda$ is decreased the probability of reject increases and probability of error decreases with diminishing returns.
}
\end{figure}

In 1970 Chow \cite{Chow70} showed a particularly simple relationship between the error probabilities and the minimum risk under the optimal decision rule
\begin{subequations}\label{ChowRisk}
\begin{align}
\mathcal R^*[\phi^*] &= \Pr(E|\theta,\lambda)+ \lambda \Pr(R|\theta,\lambda),\\
&= \int_0^\lambda d\lambda' \Pr(R|\theta,\lambda',\phi).
\end{align}
\end{subequations}
Both of these expressions can be visualized graphically, see \frf{fig6}. Prior to our work the expression given in \erf{ChowRisk} (a) is one of the ways the loss function has been explained, see e.g. \cite{DreBruKor14}. It is important that the optimal decision rule and measurement angle is used otherwise the risk will generally be different to the above risk.

It turns out that $\Pr(E)$ can be derived from $\Pr(R)$ for a particular rejection threshold. Chow~\cite{Chow70} has shown that the Stieltjes integral of $\lambda$ with respect to $\Pr(R|\theta,\lambda)$ is precisely the error probability
\begin{align}\label{StieltjesPrE}
\Pr(E|\theta,\lambda)= - \int_0^\lambda  \lambda' \, d\Pr(R|\theta,\lambda').
\end{align}
As noted by Chow, this expression is suggestive of an error probability-reject probability tradeoff relation, see \frf{fig7}. If  $\Pr(R|\theta,\lambda)$ is differentiable with respect to $\lambda$ then the Stieltjes integral reduces to the Riemann integral
\begin{align}\label{RiemannPrE}
\Pr(E|\theta,\lambda)= - \int_0^\lambda  \lambda' \left [\frac {d}{d\lambda'}\Pr(R|\theta,\lambda') \right ]d\lambda'.
\end{align}
From \erf{StieltjesPrE} and \erf{RiemannPrE} it is clear that the slope of the error-reject tradeoff curve in \frf{fig7} is exactly value of the rejection threshold. Consequently the tradeoff is most effective initially and is less rewarding as the desired errror decreases. In \frf{fig7} we also see that specifying a particular rejection threshold, e.g. $\Pr(R)=Q$ as in \cite{BagMunOli12}, implies a value for $\lambda$ and $\Pr(E)$ (once $\theta$ is fixed).

\begin{figure}\centering
  \includegraphics[width=\columnwidth]{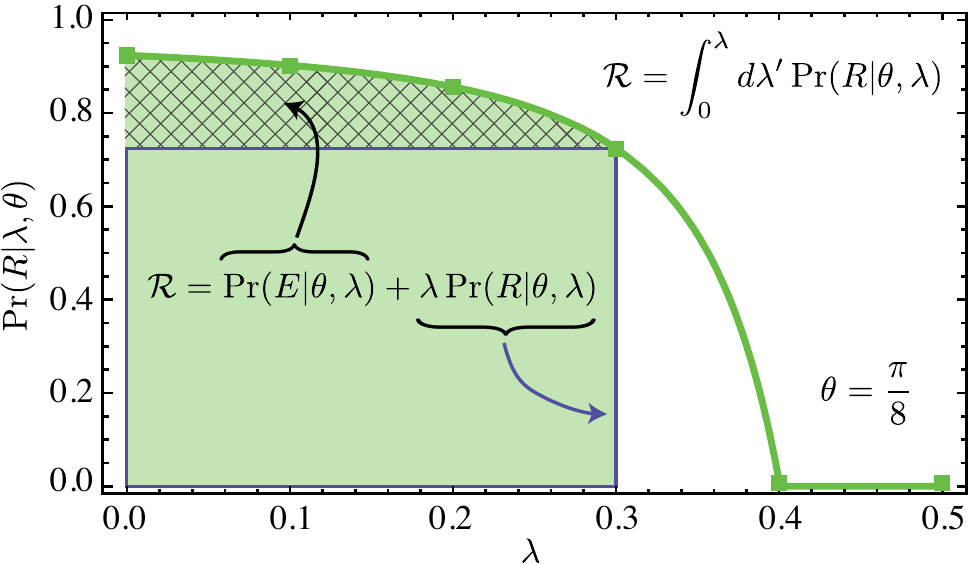}
  \caption{\label{fig6} The relationship between risk and probability for rejection. The rejection probability is plotted as a function of the rejection threshold $\lambda$ when $\theta = \pi/8$. Consider a rejection threshold of $\lambda =0.3$, given this threshold and the angle between the states the expected risk can be computed from \erf{risk_eg} to be $\mathcal R \approx 0.26$. Equation \ref{ChowRisk} (b) shows this equivalent to the (shaded) area under the curve up to the rejection threshold. The area under the curve can be decomposed into a rectangle with height $ \Pr(R|\theta,\lambda)\approx0.724 $ and width $\lambda = 0.3$ so $\lambda \Pr(R|\theta,\lambda)\approx0.2172$ the integral given \erf{RiemannPrE} results in $\Pr(E|\theta,\lambda)\approx0.0428$ and thus $\mathcal R^*= \Pr(E|\theta,\lambda)+ \lambda \Pr(R|\theta,\lambda)$.
}
\end{figure}

\begin{figure}\centering
  \includegraphics[width=\columnwidth]{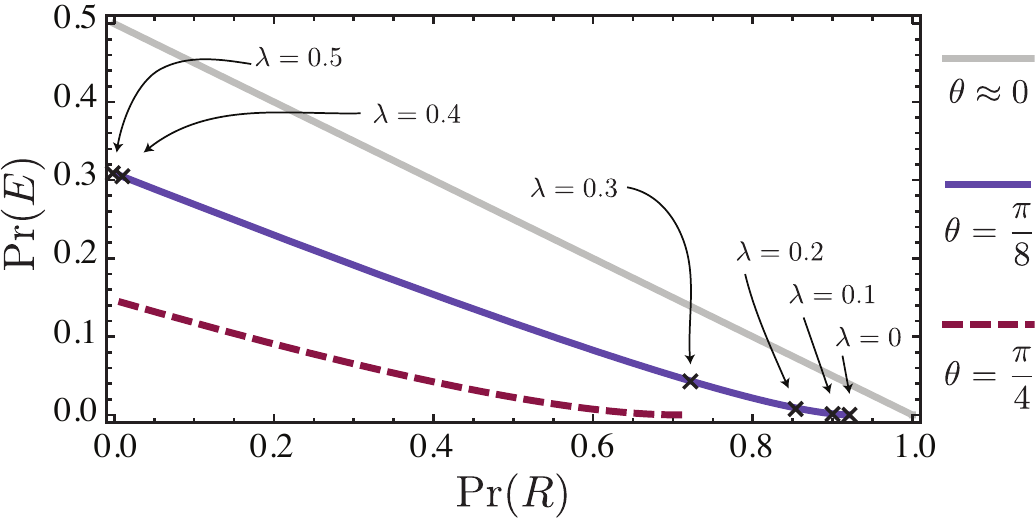}
  \caption{\label{fig7} Error-reject tradeoff curve. In fact the derivative of $\Pr(E)$ with respect to $\Pr(R)$ is $\lambda$. These curves are implicit functions of $\lambda$. The trade off is not linear in the rejection threshold $\lambda$. This is evident on the line corresponding to $\theta = \pi/8$ where six crosses corresponding to $\lambda\in [0,0.1,0.2,0.3,0.4,0.5]$ are plotted.
}
\end{figure}

\section{The 0-$\lambda_E$-$\lambda_R$ loss function}\label{beyondzol}
\begin{figure}\centering
  \includegraphics[width=0.95\columnwidth]{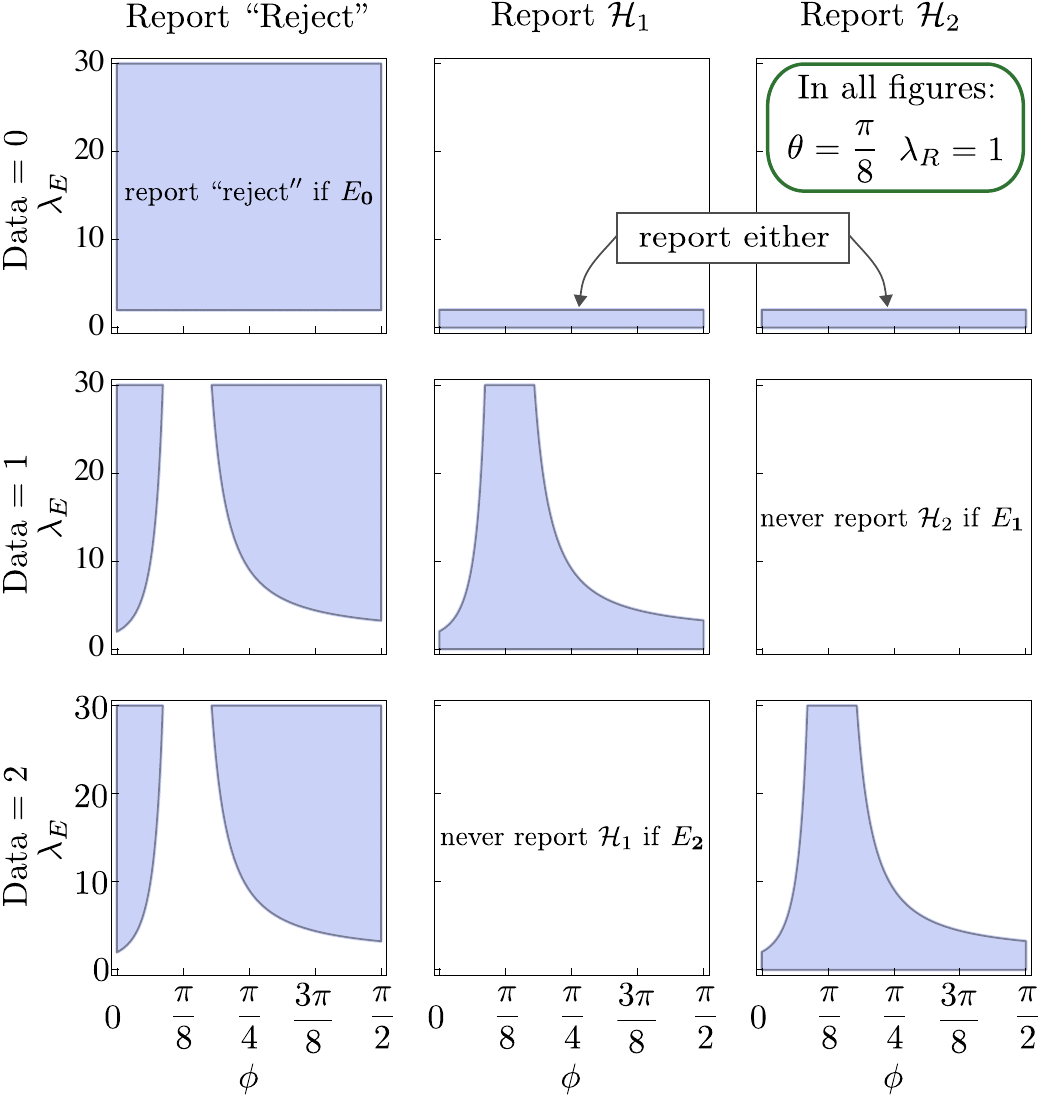}
  \caption{\label{fig8} Decision regions for the 0-$\lambda_E$-$\lambda_R$ loss function. In all figures the angle between the states is $\theta = \pi /8$ and the reject loss was chosen to be $\lambda_R=1$. The shaded regions should be intepreted as report the column heading. In row one, the reporting of a hypothesis given the inconclusinve outcome is a result of \erf{gen_risk_D0}. Evidently, as $\lambda_E$ becomes large the decision rule becomes more like unambigous state discrimination.
}
\end{figure}

Here we generalize the \zol\ loss function to the 0-$\lambda_E$-$\lambda_R$ loss function, where $\lambda_E$ is the cost of reporting the incorrect decision and $\lambda_R$ is the cost of reporting reject --i.e.,
\begin{align}
\begin{array}{rl}\label{0lamElamR}
\lambda_{1,1} &=\lambda_{2,2} = 0,\\
\lambda_{1,2} &=\lambda_{2,1} =\lambda_E,\\
\lambda_{0,1}&=\lambda_{0,2}=\lambda_R.
\end{array}
\end{align}
For the 0-$\lambda_E$-$\lambda_R$ loss function in \erf{0lamElamR} the conditional risks for decisions $i$ are
 \begin{align}\label{gen_risk}
 \begin{array}{rl}
\mathcal R [2|\mathbf{D}] &=\lambda_E [1-\Pr( \mathcal H_2| \mathbf{D} ) ],\\
\mathcal R [1|\mathbf{D}]&=\lambda_E [1-\Pr( \mathcal H_1| \mathbf{D} )  ] ,  \\
\mathcal R [0|\mathbf{D}]&= \lambda_R   ,
\end{array}
\end{align}
The following analysis assumes the same states [\erf{quantdis}], prior probablities [$\Pr(\mathcal H_i)=1/2$], and measurements [\erf{POVM}], as before. Of particular interest is the case when the measurement outcome $E_{0}(\phi)$ is obtained, i.e. $\mathbf{D}=0$, then the conditional risks are
 \begin{align}\label{gen_risk_D0}
 \begin{array}{rl}
\mathcal R [2|0] 
&=\lambda_E/2,\\
\mathcal R [1|0]
&=\lambda_E  /2, \\
\mathcal R [0|0] 
&= \lambda_R. 
\end{array}
\end{align}
Thus if  $\lambda_R> \lambda_E/2$ we should never reject, instead we should report either hypothesis, as illustrated in row 1 of \frf{fig8}. In \frf{fig8} we have  chosen $\lambda_R=1$ so that for all $\lambda_E\le 2$ we must report either hypothesis to minimize our risk. In particular if  we perform the a measurement with an inconclusive outcome $\phi<\pi /2$ and obtain the inconlusive outcome $E_{0}$ we should randomly choose between reporting $\mathcal H_1$ and $\mathcal H_2$. For $\lambda_R < \lambda_E /2$ we find 
\begin{align}\label{gen_decision}
\delta(\mathbf{D}) =  \begin{cases}
2 & \text{if }\Pr( \mathcal H_2| \mathbf{D} )\ge 1- \frac{\lambda_R}{\lambda_E} \text{ and }
   \Pr( \mathcal H_1| \mathbf{D} ) \\
1 & \text{if }\Pr( \mathcal H_1| \mathbf{D} )\ge 1-  \frac{\lambda_R}{\lambda_E} \text{ and } 
 \Pr( \mathcal H_2| \mathbf{D} ) \\
0 & \text{otherwise} 
\end{cases}.
\end{align}
In words, the decision rule is as follows: find the largest posterior probability; if it is greater than or equal to the threshold $1- \frac{\lambda_R}{\lambda_E} $, report it; if it is less than $1- \frac{\lambda_R}{\lambda_E}$, report ``reject''. 

Like the \zol\ loss function, the 0-$\lambda_E$-$\lambda_R$ loss function also interpolates between the Helstrom measurement and unambiguous state discrimination, as illustrated in \frf{fig9}. Notice, for both loss functions, we did not need to ``normalize'' the loss function or add additional contraints such as $\Pr(R)=0$ or $\Pr(E)=0$, unlike other approaches \cite{DreBruKor14}.

\begin{figure}\centering
  \includegraphics[width=0.95\columnwidth]{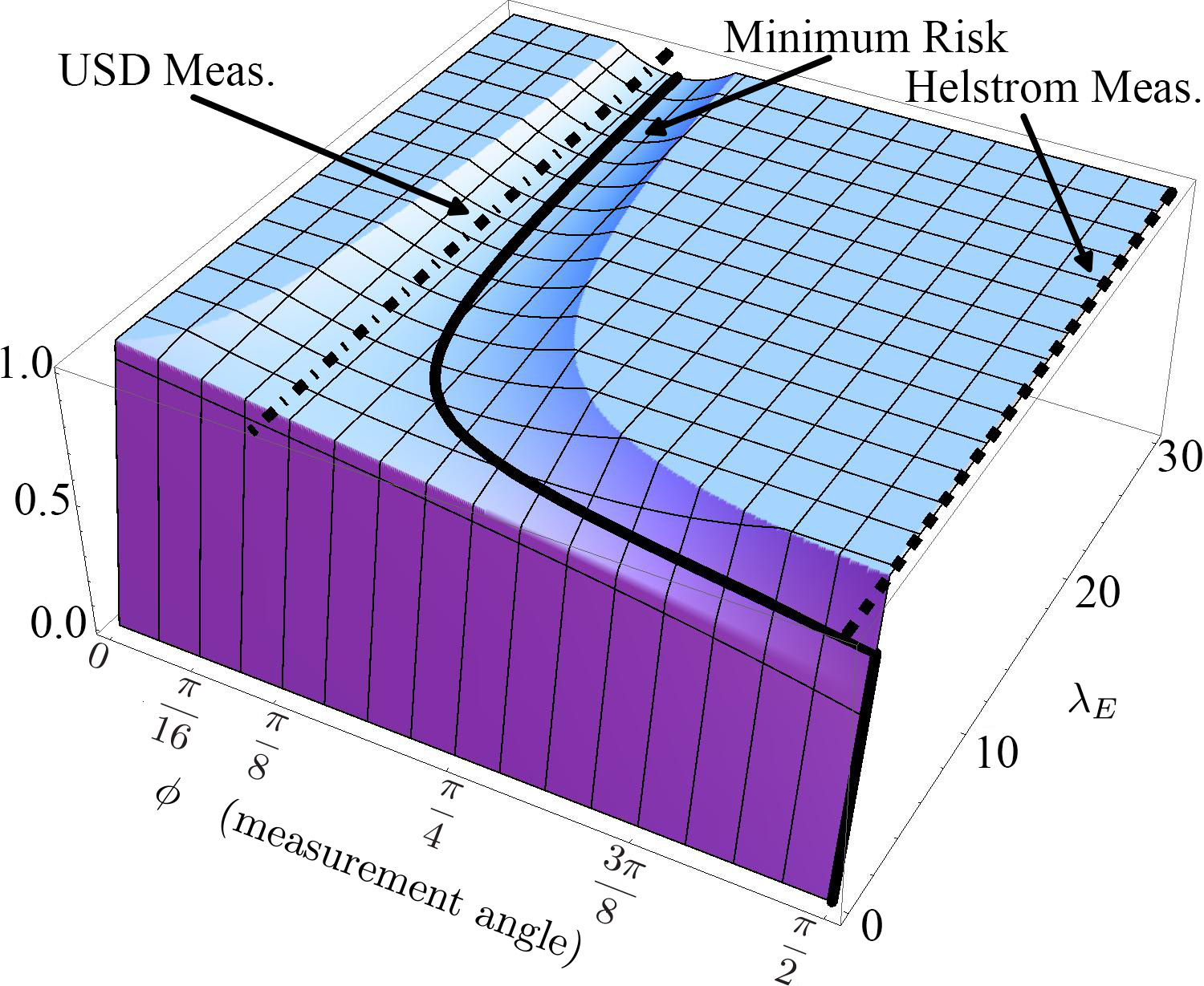}
  \caption{\label{fig9} Risk as a function of measurement angle $\phi$ and the cost of reporting the wrong decision $\lambda_E$ for the 0-$\lambda_E$-$\lambda_R$ loss function. Here $\theta = \pi /8$ and the reject loss was chosen to be $\lambda_R=1$. For $\lambda_E<2.5$ we see the optimal measurement is the Helstrom measurement and as $\lambda_E\rightarrow \infty$ the optimal measurement approaches the USD measurement.
}
\end{figure}

\section{discussion}
In the ongoing debate about postselection for information theoretic tasks in quantum theory, we have given a plausible example where postselection is a feature of the optimal solution.  We say plausible because the loss function on the decisions was not tailored to favor full-blown postselection---the solution was not obvious.  

In \srf{sec:statediscrim} we have shown that USD measurements only arise in the limit when the cost assigned to discarding data is exactly zero, which corresponds to the line $\lambda=0$ for all $\theta$ in \frf{fig3}.  In contrast, the Helstrom measurement appears to be the natural measurement for approximately half of the paramter space $\lambda \gtrsim \half (1-\theta/2)$. For the remainder of the parameter space, i.e. $\lambda \lesssim \half (1-\theta/2)$, strategies involving postselection (that are not USD) are optimal. In \srf{sec:pr_pe} we unified three seemingly separate approaches, namely the decision theoretic approach (i.e. our \zol\ loss function), the rejection threshold approach \cite{BagMunOli12}, and the probability tradeoff approach \cite{DreBruKor14}.  Section \ref{beyondzol} highlighted that the decision function can not simply be ignored---in some situations it is better to report an answer even if the inconclusive outcome was obtained.

It is natural to ask what the implications of our analysis are. In practical situations it could be desirable to reduce errors by rejecting some data, but excessive rejection is required to reduce error to zero. And, at the point where the error is zero one can equivalently reject without bothering to perform any experiment, as the cost of rejection is also zero. Generally this implies when a loss function is specified as conditional on some event being successful that this is equivalent to assigning {\em cost} to a rejection option. Again, if the cost of rejection is zero why should you bother to perform the experiment at all? We have suggested a sensible approach is to embed a postselection protocol into a class of protocols which assign loss for discarding data, this makes clear the price of postselection. 

For example, consider offline magic state distillation for quantum computation \cite{BravyiKitaev05}. The success probability is relevant for quantifying efficiency (or expected yield in Sec. VI. of \cite{CamAnwBro12}) of the magic state distillation routine. When the success probability for the scheme is too small then the overall distillation routine is inefficient, even if it performs very well when it does succeed. This is generically true in offline state preparation. If costs are low, we are happy to  wait for some time for a state to be prepared. But the cost are not zero, as we actually want to make a state and perform a useful task.

The virtue of the decision theoretic approach is that all the assumptions, constraints and figures of merit are made explicit at the outset---the rest is derived. Thus, within this framework it is quite natural to include new constraints and features. For example, if experimental noise or inaccuracies or constraints are of concern, one must include those at the highest level---that is, they must be specified in the initial states, POVM, or loss function. Questions of robustness or imperfections, which plague other approaches, are simply a category mistake to ask here.

A number of open questions remain. The first class of questions are about extensions to the specific ideas developed in this manuscript. A simple modification is when Alice makes collective measurements on $N$ copies of $\ket{\Psi_1}$ or $\ket{\Psi_2}$. In this case the states look more orthogonal because $|\ip{\Psi_1}{\Psi_2}|^{2N}\le |\ip{\Psi_1}{\Psi_2}|^2$. Based on our results in \frf{fig3} we conjecture that the optimal joint measurement for the \zol\ loss function will look closer to a Helstrom measurement than the USD measurement.   The obvious question is: does a bound on the $N$ copy risk exist? Ideally the solution would be something like the quantum Chernoff bound~\cite{QChernoff} which bounds the minimum error probability asymptotically in $N$ (i.e. the risk of the 0/1 loss function).

The second class of questions are about the role of postselection in quantum information tasks. Although we have conjured an exotic loss function for which the optimal strategy includes postselection, it is not tied explicitly to an existing operational task. 
Nevertheless we suggest that our decision theoretic approach should be taken for any practical state discrimination (or estimation) problem which allows for the possibility of postselection. Extending our approach to parameter estimation seems to be the next great challenge. The results in this manuscript add weight to our suggested loss function \cite{ComFerJia13a}: report ``reject" and incur loss $\lambda$ for mean squared error (MSE) above some threshold and incur the MSE loss below that threshold.

\acknowledgments{We thank Emili Bagan, Ben Baragiola, John Calsamiglia, Carl Caves, Justin Dressel, Bernat Gendra, Chris Granade, Mark Howard, Norbert L\"utkenhaus, Yihui Quek, Ramon Mu\~{n}oz-Tapia, and Elie Wolfe for discussions and suggestions. We are particularly grateful that Elie pointed out \erf{phi_opt} could be simplified to its present form. The authors thank Mathematica-gicians Agata Bra\'nczyk and Chris Granade---without the magic the figures in this manuscript would look considerably different. This work was supported in part by NSF Grant Nos. PHY-1212445 and PHY-1314763. JC was also supported by the Australian Research Council Centre of Excellence for Engineered Quantum Systems grant number CE110001013,  CERC, NSERC, and FXQI. CF was also supported in part by the Canadian Government through the NSERC PDF program, the IARPA MQCO program, the ARC via EQuS project number CE11001013, and by the US Army Research Office grant numbers W911NF-14-1-0098 and W911NF-14-1-0103.}


\begin{thebibliography}{99}
\bibitem{explicit}
Some examples are
\begin{align*}
\Pr(E_{\mathbf 2}|\mathcal H_1,\phi) =\!\frac{1-\cos (\theta -\phi )}{2[1+ \cos (\phi )]} ,\,\,
\Pr(E_{\mathbf 1}|\mathcal H_1,\phi) =\!\frac{1-\cos (\theta +\phi )}{2[1+ \cos (\phi )]},\\
\text{and }\Pr(E_{\mathbf 0}|\mathcal H_1,\phi) =\cos ^2(\theta /2) \cos (\phi ) \sec ^2(\phi /2),\quad\quad\quad
\end{align*}
which infact specify all the probablities due to the symmetry assumptions in \erf{symm}.
\end{thebibliography}
\end{document}